\documentclass[12pt,a4paper]{article}
\usepackage{authblk}
\usepackage{amsmath}
\usepackage{amsfonts}
\usepackage{amssymb,float}
\usepackage{latexsym}
\usepackage[latin1]{inputenc}
\usepackage{amssymb,float}
\usepackage{latexsym}
\usepackage{epstopdf}
\usepackage{geometry}
\usepackage[active]{srcltx}
\usepackage{graphicx}
\usepackage{epstopdf}
\usepackage{wrapfig}
\usepackage{caption}
\usepackage[T1]{fontenc}
\usepackage{subcaption}
\usepackage{enumerate}
\usepackage[ bookmarks=true,         bookmarksnumbered=true, colorlinks=true,
pdfstartview=FitV, linkcolor=blue, citecolor=blue, urlcolor=blue]{hyperref}
\usepackage{amssymb}%
\providecommand{\U}[1]{\protect\rule{.1in}{.1in}}
\begin{document}
\title{\bf ARMA Model Development and Analysis for Global Temperature Uncertainty}
\author[1]{Mahmud Hasan\thanks{mhasan4@go.olemiss.edu}}
\author[1]{Gauree Wathodkar\thanks{gkwathod@go.olemiss.edu}}
\author[1]{Mathias Muia\thanks{mnmuia@go.olemiss.edu}}
\affil[1]{Department of Mathematics, University of Mississippi, Oxford, MS, USA}

\maketitle

	\begin{abstract}
 Temperature uncertainty models for land and sea surfaces can be developed based on statistical methods. In this paper we developed a novel time series temperature uncertainty model 
 which is the Auto-regressive Moving Average  (ARMA)$(1,1)$ model. The model was developed for observed annual mean temperature anomaly $X(t)$ which is a combination of true (latent) global anomaly $Y(t)$ for a year $(t)$ and normal variable $w(t)$. The uncertainty is taken as the variance of $w(t)$ which was decomposed to Land Surface Temperature (LST) uncertainty, Sea Surface Temperature (SST) uncertainty, and the corresponding source of uncertainty. The ARMA model was analyzed and compared with Auto-regressive (AR), and Auto-regressive integrated moving average (ARIMA) for the data taken from NASA, Goddard Institute for space studies Surface Temperature Analysis. The statistical analysis of the Auto-correlation function (ACF), Partial auto-correlation function (PACF), Normal quantile-quantile (Normal Q-Q) plot, the density of the residuals, and variance of normal variable $w(t)$ shows that ARMA$(1,1)$ fits better than AR$(1)$ and ARIMA$(1,d,1)$ for $d=1,2$.
	\end{abstract}
 

	\section{Introduction}
Temperature uncertainty can have a significant impact on astronomical research in several ways.
Observations made using telescopes and other astronomical instruments are often temperature sensitive. As the temperature changes, so do the sensitivity and response of the instrument, leading to measurement errors if the temperature is not accurately monitored and corrected. The quality of astronomical data is also affected by temperature variations. For example, fluctuations in the temperature of the detectors used to observe light from stars can introduce noise into the data that is difficult to distinguish from true signals.
Therefore, temperature control and accurate temperature measurements are important considerations in many areas of astronomical research, and researchers often go to great lengths to minimize the impact of temperature uncertainty on their results.

Temperature uncertainty in sea surface temperature (SST) \cite{Ra} and land surface temperature (LST) \cite{Qu} is typically modeled using statistical methods. One common time series model for temperature uncertainty is the Auto-regressive Moving Average (ARMA) \cite{Ro} model. In this model, the temperature at a given time point is modeled as a function of its past values and the residuals (errors) from previous time points. However, it is possible developing an ARMA model based on sea surface temperature (SST) and land surface temperature (LST) uncertainty. There is other time series model like Auto-regressive, Auto-regressive integrated moving average(ARIMA) \cite{Ro}, Seasonal Auto-regressive integrated moving average (SARIMA) \cite{Ro} and Generalized Auto-Regressive Conditional Heteroskedasticity (GARCH) \cite{Ro} which can be used for temperature uncertainty data.

Land surface temperatures are available from the Global Historical Climate Network-Monthly (GHCNm) \cite{Me}. Sea surface temperatures are determined using the extended reconstructed sea surface temperature (ERSST) \cite{Hu} analysis. ERSST [15] uses the most recently available International Comprehensive Ocean-Atmosphere Data Set (ICOADS) \cite{Fr} and statistical methods like ARMA and ARIMA  that allow stable reconstruction using sparse data. 

James Hansen defined the GISS temperature analysis scheme in the late 1970s when a method of estimating global temperature change models needed for comparison with one-dimensional global climate models. The analysis method was fully documented by \cite{Han}. The analysis sub-sampled a long run of the GISS-ER \cite{Han2} climate model according to the periods of the station network on the Earth during these three time periods. Another sophisticated uncertainty model based on Global and Regional average temperature anomaly time series analysis was developed by \cite{Rob}. Very Recently, \cite{Mor1} made an interpolation approach to generate a Kriging-based field using an assumed distance-based optimization technique. In this paper, we are developing a statistical uncertainty model inspired by \cite{Len} that shows better than some other models. However, Optimizing the data for the model need to be explored which could be future research by the corresponding optimization technique \cite{Mah2},\cite{Mah3}. 

The uncertainty models are based on existing methods and give predictions of temperature uncertainty. Those models are the improvement of uncertainty analysis for the GISTEMP based on the estimation of the probability for the previous year's data. In this paper, we develop a Temperature uncertainty model which is ARMA $(1,1)$. If we consider, the annual mean
temperature anomaly $X(t)$ as a linear combination of true (latent) global anomaly $Y(t)$
for time(year) (t) and random variable $w(t)\sim N(0,\sigma^2)$. The uncertainty is defined as the variance of $w(t)$ that can be decomposed to Land Surface Temperature (LST) uncertainty, Sea Surface Temperature (SST) uncertainty, and the corresponding source of uncertainty. Moreover, a difference series derived by $X(t)$ and $Y(t)$ which gives ARMA model after introducing systematic bias.\\

The ARMA model was justified by comparing with AR and ARIMA models using some time series property Auto-correlation Function (ACF), Partial autocorrelation function (PACF), and density residual. We added a new property uncertainty which measures the fitness of the model for the corresponding data. For analyzing the model, we are taking data from NASA, Goddard Institute for space studies Surface Temperature Analysis is the source of comprehensive global surface temperature data set spanning 1880 to the present at monthly resolution. The model was for the corresponding data using the autoarima function in python.
We organized the paper into different sections.

In section $2$, we developed the ARMA model for the observed annual mean temperature anomaly $X(t)$ at time $t$. The variable $X(t)$ was decomposed for true (latent) global anomaly $Y(t)$ of temperature for a year $t$ and normal variable $w(t)$. Then normal variable decomposed to Land temperature anomaly and sea temperature anomaly with a true anomaly at $t-1$ equal to the observed mean temperature anomaly $X(t)$.  We used the data using the autoarima function in python for all the models.

In section $3$, after developing the model we find the proposed model is ARMA$(1,1)$ gives the scope of discussing attributes like Auto-correlation function (ACF), Partial auto-correlation function (PACF), Normal quantile-quantile (Normal Q-Q) plot, the density of residuals
affects the ARMA model. The variance for the variable $w(t)$ affects the model which was explained in the data analysis and discussion section. Each attribute was explained by mathematical evaluation and the role of the corresponding parameter.

In section $4$, a detailed analysis and discussion of our results were done. Firstly, the non-stationarity of GISTEMP data was verified using the Augmented Dickey-Fuller Test (ADF test). Then there is detailed analysis for AR$(1)$, ARMA$(1,1)$, ARIMA$(1,1,1)$ and ARIMA$(1,2,0)$ models with the diagnosis and residual analysis. The comparison of the models was explained based on parameter estimation. We used the data using the autoarima function in python for all the models.

The last section concludes the results we found through the theoretical finding and the numerical analysis.

 \subsection{ Preliminary and definition of model}
 We can get the corresponding definition and notations for the model from \cite{Ro}. In this section, we are going to discuss the preliminary definition and corresponding coefficient parameter of the model. AR, ARMA and ARIMA.
 \subsubsection*{Auto-regressive (AR)}
	An auto-regressive \textbf{AR(p)} model of order $p$, for the current value for time $t$
 \begin{align}
     X_{t}=\phi_{1}X_{t-1}+\phi_{2}X_{t-2}+\phi_{3}X_{t-3}+..........+\phi_{p}X_{t-p}+w_{t}
 \end{align}
 
  $X_{t}$ is stationary and parameters  $\phi_{1},\phi_{2},\phi_{3},..........\phi_{p}$  are constants with $\phi_{p}\not= 0$ and $w_{t}\sim WN(0,\sigma_{w}^2)$. By backshift operator \textbf{AR(p)} can be written as 
 \begin{align}
     \phi(B)X_{t}=\epsilon_{t}
 \end{align}
 where $\phi(B)=(1-\phi_{1}B-\phi_{2}B^2-......-\phi_{p}B^p)$. The ACF of $AR(1)$ is 
 \begin{align}
     \rho(h)=\frac{\gamma(h)}{\gamma(0)}=\phi^{h}, \ h\geq 0
 \end{align}
 where auto covariance function $\rho(h)$ satisfies $\rho(h)=\phi\psi(h-1),\ h=1,2,\cdots.$

\subsubsection*{Auto-regressive moving average (ARMA) }
A time series as ${x_{t}; t = 0, ±1, ±2, . . .}$ is ARMA$(p, q)$ if it is stationary and
\begin{align}
  X_{t}=\phi_{1}X_{t-1}+\phi_{2}X_{t-2}+\phi_{3}X_{t-3}+..........+\phi_{p}X_{t-p}
  + \theta_{1}w_{t-1}+\theta_{2}w_{t-2}+\theta_{3}w_{t-3}+..........+\theta_{q}w_{t-q}
\end{align}
where $\phi_{p},\theta_{q}\neq 0$ and $\sigma_{w}^2\geq 0$. The parameters $p$ and $q$ are called the auto-regressive and the moving average orders, respectively.

\subsubsection*{ Auto-regressive integrated moving average (ARIMA)  }
A process $X_{t}$ is said to be ARIMA$(p, d, q)$ if
\begin{align}
    \nabla^d X_{t} =(1-B)^d X_{t}
\end{align}

is ARMA$(p, q)$for the seasonality parameter $d$ and back-shift parameter $B$. In general, we will write the model as
\begin{align}
    \phi(B)(1-B)^d X_{t}=\theta(B) w_{t}.
\end{align}
The coefficients of an AR, ARMA, and ARIMA model play a crucial role in determining the model's ability to capture the behavior of a time series, and their choice can have a significant impact on the model's predictions. In our model, the coefficient parameter has been introduced as the variance of temperature which affects the comparability of the model with others attributes.

\section{Uncertainty ARMA Model} 
	Let $Y(t)$ be the true (latent) global anomaly of temperature for a
year t, we view the calculated (the observed) annual mean temperature anomaly
\begin{equation}\label{X}
		X(t) = Y(t)+w(t).
	\end{equation}
 The random variable $w(t)\sim N(0,\sigma^2)$. The uncertainty in our calculation of the global
mean anomaly is then defined as
\begin{equation}
	W(t)=\sigma^2.
	\end{equation}
We can decompose  total uncertainty as

\begin{equation}\label{W}
W(t)=\sigma ^2=\sigma_{L} ^2 + \sigma_{S} ^2.
	\end{equation}
 Where uncertainty decomposed into two components: the uncertainty in the global mean anomaly
due to uncertainties in the land calculation  $\sigma_{L} ^2$  and uncertainty in the global mean anomaly due to uncertainties in the sea surface calculation  $\sigma_{S} ^2$.
It means there must exist random variable for anomaly due to uncertainties in the land $w_{L}(t)\sim N(0,\sigma_{L}^2)$ and anomaly due to uncertainties in the sea surface $w_{S}(t)\sim N(0,\sigma_{S}^2)$ for which we can write (1) as 
\begin{equation}\label{X2}
		X(t) = Y(t)+w_{L}(t)+ w_{S}(t).
	\end{equation}
	
Reduced coverage global annual means, $X_{i}(t)$, are calculated for each of the 14 decadal time periods using a modified GISTEMP procedure. Where $i$ represents the decade used and $t$ represents the time in year . The difference series for decade $i$ is 

\begin{equation}
		D_{i}(t)= Y(t)-X_{i}=w_{iL}(t)+ w_{iS}(t).
	\end{equation}
	
We introduce a potential systematic additive bias $\alpha_{i}$ and multiplicative
bias $\beta _{i} $. Then (1) can be formulated as 

\begin{equation}\label{X3}
		X_{i}(t)= \alpha_{i}+ \beta _{i} Y(t)+   w_{iL}(t)+ w_{iS}(t).
	\end{equation}
 Then decomposing Land temperature anomaly and sea temperature anomaly with true anomaly for $t-1$ equal to observed mean temperature anomaly $X(t)$ . Then we can write 
	\begin{equation}\label{X4}
		X_{i}(t)= \alpha_{i}+ \beta _{i} X(t-1)+   w_{iL}(t)+ w_{iS}(t-1).
	\end{equation}
\textbf{Remark}

The equation \eqref{X4} represents the ARMA $(p,q)$ model for $p=1,q=1$ where $X_{i}(t)= \alpha_{i}+ \beta _{i} X(t-1)$ is Auto-regressive of order one i.e AR$(1)$ and $X_{i}(t)= w_{iL}(t)+ w_{iS}(t-1)$ is Moving Average of order one i.e MA$(1)$. The  coefficients $w_{iL}$ and $ w_{iS}$  of MA$(1)$ represents land surface and sea surface temperature uncertainty. By taking data from NASA, Goddard Institute for space studies Surface Temperature Analysis we show that this coefficient affects ARMA model \eqref{X4} to fit better than AR$(1)$ and ARIMA. Besides, coefficients  we also found that other time series property ACF, PACF, Normal quantile-quantile (Normal Q-Q) plot, the density of residuals for \eqref{X4} fits better.
 
\section{Statistical Characteristics of ARMA(1,1)}
In this section, we are going to explain time series property that justifies fitting the better model. In our case, the property  Auto-correlation function (ACF), Partial auto-correlation function (PACF), Normal quantile-quantile (Normal Q-Q) plot, the density of residuals affects the ARMA model. Another property is the variance for the variable $w(t)$ affecting the model that are discussed in the data analysis section. 

\subsection{ Auto-correlation Function (ACF)}	

If we write ARMA(1,1)  as which is casual form represent $X_{t}=\sum_{j=1}^{\infty}\psi_{j}w_{t-j}$, \begin{equation*}
	    X_{t}=w_{iL}(t)+(1+\beta_{i})w_{iS}(t-1),\ \text{for}\ 0<\beta_{i}<1,
	\end{equation*} then 
\begin{equation*}
	    X_{t}=w_{iL}(t)+(1+\beta_{i})w_{iS}(t-1).
	\end{equation*}	
 The corresponding  auto correlation (ACF) from \cite{Ro}  can be given as 
\begin{equation}\label{rh}
         \rho(h)=\frac{(1+\beta_{i})^{2}}{2(1-\beta_{i})}\beta_{i}^{h-1}\\
        =\frac{1}{2}(1+\beta_{i})\beta_{i}^{h-1}.
\end{equation}
In equation \eqref{rh}, $h$ represents the time difference, and $\beta_{i}$ are parameter coefficients. Depending on the parameter value we have to justify the fitness of the model. The auto-correlation function (ACF) defines how data points in a time difference i.e lag are related, on average, to the preceding data points.

\subsection{ Partial Auto-correlation Function (PACF)}	

However AR$(1)$ and ARMA$(1,1)$ processes are fully correlated, their ACF tails off and never becomes zero, though it may be very close to zero. In such cases, sometimes it may not be possible to identify the process on the ACF basis only. So we will consider  Partial Auto-correlation Function (PACF), which together with the ACF will help to identify the models. 
The PACF of a zero-mean stationary time series $\lbrace X_{t}\rbrace_{t=0,1,2,\cdots}$ is defined as\begin{equation}
    \begin{split}
        \phi_{11}&=corr(X_{1},X_{0})=\rho(1)\\
        \phi_{\tau\tau}&=corr(X_{\tau}-f_{(\tau-1)},X_{0}-f_{(\tau-1)}),\ \tau\geq 2,
    \end{split}
\end{equation}
where 
\begin{equation*}
    f_{\tau-1}=f(X_{\tau},\cdots, X_{1}) 
    \end{equation*}
minimizes the mean square linear prediction error
    \begin{equation*}
        E(X_{\tau}-f_{(\tau-1)})^{2}.
    \end{equation*}
The subscript at the $f$ function denotes the number of variables the
function depends on. $\phi_{\tau\tau}$ is the correlation between variables $X_{t}$ and  $X_{t-\tau}$ with the linear effect. Basically, the parameter value $\phi$ estimates the fitness of ARMA model.

\subsection{Normal quantile-quantile (Normal Q-Q) plot}

A normal quantile-quantile (Q-Q) plot is a graphical method for assessing whether a set of sample data are approximately normally distributed. It compares the quantiles of the sample data to the quantiles of a theoretical normal distribution.

In the context of an ARMA (AutoRegressive Moving Average) model, a normal Q-Q plot can be used to assess the normality of the residuals, which are the differences between the observed values and the values predicted by the ARMA model. If the residuals are normally distributed, it indicates that the ARMA model has captured the majority of the systematic patterns in the data, and the remaining differences are random noise that can be well approximated by a normal distribution.

In other words, if the residuals of an ARMA model are well approximated by a normal distribution, it suggests that the model is a good fit for the data. However, if the residuals deviate significantly from normality, it may indicate that the ARMA model is not a good fit and that other modeling techniques or modifications to the ARMA model should be considered.

\subsection{Forecasting}

Here we are presenting the forecasting method for ARMA though forecasting can not be used as the property but we can get parameter estimation. The goal of forecasting is to predict future values of a time series, based on the collected present data. For the data $x_{1},x_{2}.....x_{n}$ we write forecasting model as 

\begin{equation*}
    X_{n+1}=\beta_{i}X_{n}+w_{iL}(n)+w_{iS}(n) 
\end{equation*}	
where $\beta_{i}$ is the AR parameter coefficient.
One-step ahead truncated forecast is
\begin{equation*}
    \tilde{X}_{n+1}^{n}=\beta_{i}X_{n}+0+\tilde{w}_{n}^{n},
\end{equation*}	
Using truncated forecast: $\tilde{w}_{0}^{n}=0,\ \tilde{w}_{1}^{n}=X_{1}$. Then
\begin{equation*}
    \tilde{w}_{t}^{n}=X_{t}-\beta_{i}X_{t-1}-\tilde{w}_{t-1}^{n}, t=2,\cdots, n.
\end{equation*}	
Approximate prediction \cite{Ro}
\begin{equation*}
    p_{n+m}^{n}=\gamma_{w}^{2}\left[1+(1+\beta_{i}^{2})\sum_{j=1}^{m-1}\beta_{i}^{2}(j-1)\right]
\end{equation*}
\begin{equation*}
    p_{n+m}^{n}=\gamma_{w}^{2}\left[1+(1+\beta_{i})^{2}\frac{(1-\beta_{i}^{2(m-1)})}{1-\beta_{i})^{2}}\right].
\end{equation*}
$1-\alpha$ prediction intervals are $X_{n+m}^{n}\pm C_{\alpha/2}\sqrt{p_{n+m}^{n}}$, where $C_{\alpha/2}$ is the degree of confidence.\\

When computing prediction intervals from data, we substitute estimates for parameters, giving approximate prediction intervals. The prediction interval gives the estimates of coefficient parameter $\beta_{i}$. In general, we need better estimates from truncated forecast  and it is possible to check for model stability and check forecasting ability of model by withholding data.

	\section{Numerical Data Analysis and Discussion }
 \begin{figure}[h]
		\centering
  \rotatebox{90}{\textcolor{black}{ \small{Average global temperature anomaly}}}
		\includegraphics[scale=0.35]{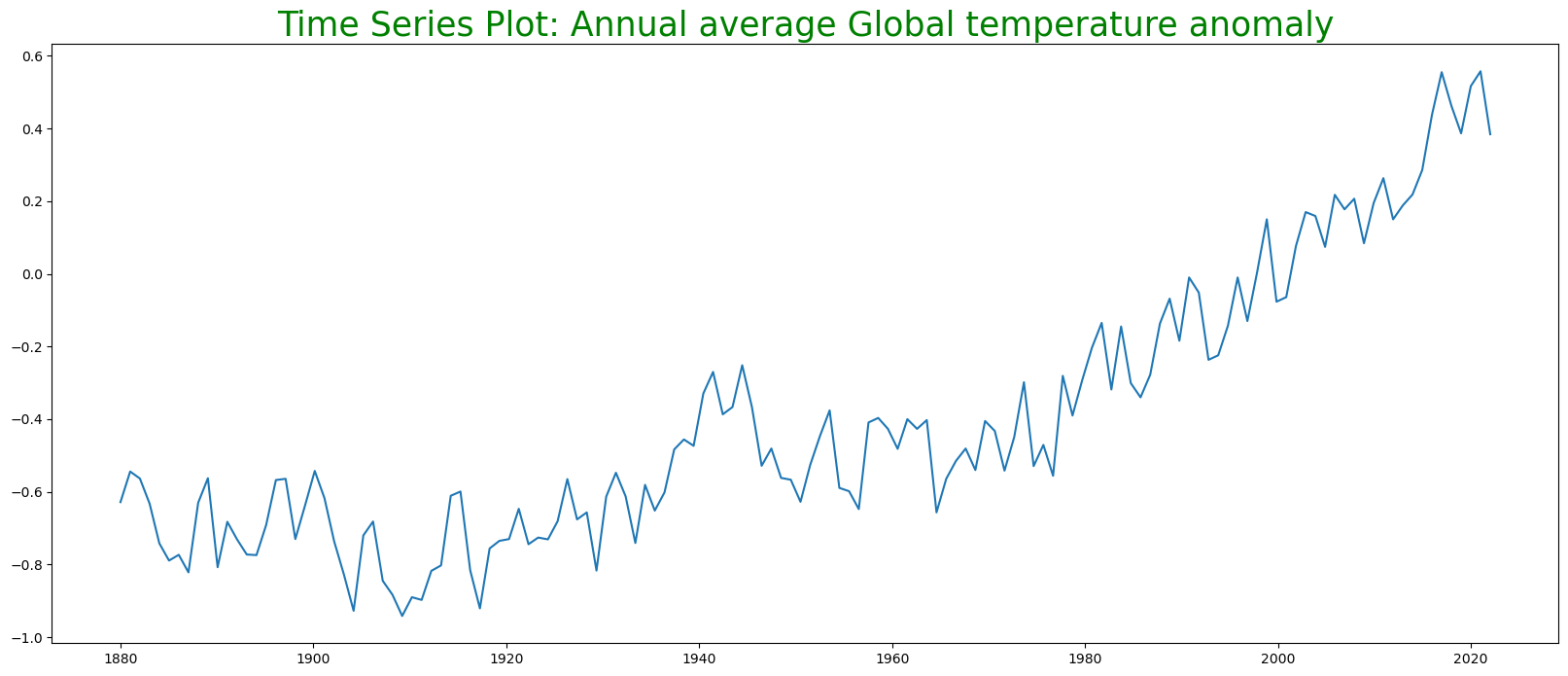}
  
  \vspace*{0.1cm}\hspace*{2.0cm}\textcolor{black}{year}
		\caption{Time series of Global temperature anomaly data}
		\label{fig1}
	\end{figure}
	\paragraph{}We will consider the annual Global temperature anomaly data from NASA, Goddard Institute for Space Studies Surface Temperature (GISTEMP) temperature Data since $1880$. 
 
    In 2019, Lenssen, N., G. Schmidt, J. Hansen, M. Menne, A. Persin, R. Ruedy, and D. Zyss discussed about different aspects of GISTEMP data in \cite{Len}. There they discussed about confidence intervals of annual mean of the data and also gave confidence intervals for Ocean temperature anomaly. Further they suggested that AR$(1)$, can be a reasonable model for comparing this data for short time periods. In this article, we try to do the detailed analysis for AR$(1)$ model and extend this discussion further to the complex models like ARMA$(1,1)$ and ARIMA$(1,d,1)$ for $d=1,2$.

\begin{figure}[h]
     \centering
     \begin{subfigure}[b]{0.45\textwidth}
         \centering
         \includegraphics[width=\textwidth]{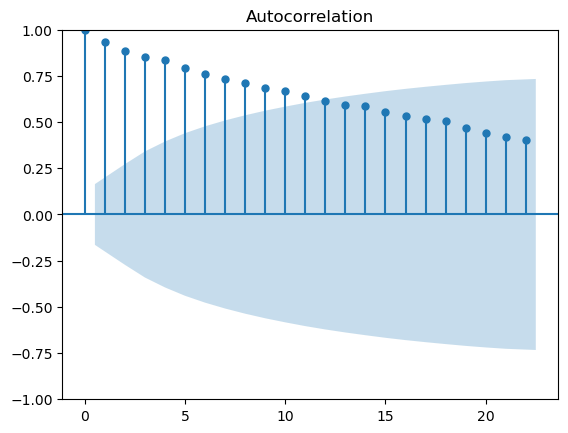}
          \vspace*{0.1cm}\hspace*{2.0cm}\textcolor{black}{lags}
         \caption{Autocorrelation for Global temperature data}
         \label{fig2}
     \end{subfigure}
     \hfill
     \begin{subfigure}[b]{0.45\textwidth}
         \centering
         \includegraphics[width=\textwidth]{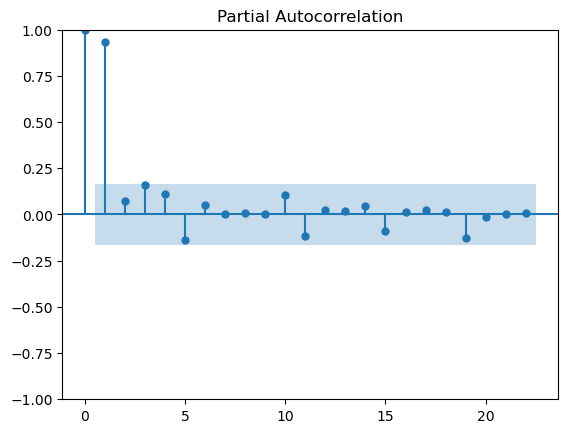}
         \vspace*{0.1cm}\hspace*{2.0cm}\textcolor{black}{lags}
         \caption{Partial Autocorrelation }
         \label{fig3}
     \end{subfigure}
     \hfill
        \caption{ACF and PACF for Global temperature data}
\end{figure}

 \paragraph{}Figure \ref{fig1} shows the time series plot for Annual average Global temperature anomaly. Since we are considering the annual average temperature anomalies, this graph does not have seasonality. Here we can see that earlier the Global temperature anomaly had a trend which was oscillating about some mean till $1960$. But since $1960$ there is a clear uptrend in the global temperature anomaly graph. It has increased in significant amounts and thus, predicting some future data for global temperature anomaly is very important.

	\paragraph{}Figure \ref{fig2} indicates the auto-correlation function for the Global temperature anomaly time series. This are the auto-correlation values for the first $20$ lags. Here the shaded region shows the threshold. In this graph, we can see that the auto-correlation decay slowly. So, there is a possibility that the time series is not stationary as the spikes of the ACF plot are over the threshold region.

	\paragraph{}Figure \ref{fig3} shows the Partial auto-correlation (PACF) for the time series for global temperature anomaly. Here, First two lags are outside the threshold and then PACF shows a sudden drop such that all other lags have PACF inside the threshold.

	\paragraph{}Using Augmented Dickey Fuller Test (ADF test) \cite{ADF} we will check if the time series is stationary or not. For this python package statsmodels.tsa.stattools was used. The result for the ADF test was generated as follows:

	\includegraphics[scale=0.5]{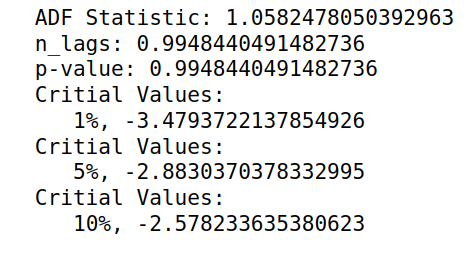}

	When we perform an ADF Test on the data, the p-value obtained is greater than the significance level of 0.05 and the ADF statistic is higher than any of the critical values. That means, there is no reason to reject the null hypothesis. So, the time series is in fact non-stationary.

        \begin{figure}[h]
		\centering
		\includegraphics[scale=0.9]{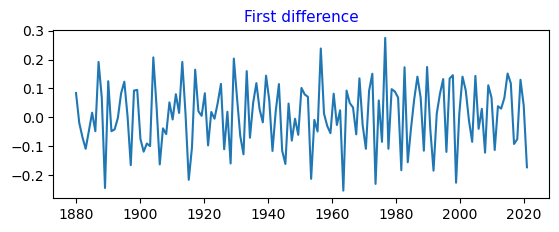}
              
            \vspace*{-0.1cm}\centering{\textcolor{black}{Year}}
  		\caption{First difference time series of the Global temperature anomaly data}
		\label{fig4a}
	\end{figure}
 
	\paragraph{}Thus we try to find a stationary time series. For that we take the first difference of the given data. Figure \ref{fig4a} shows the first difference of the given time series. It is certainly not up-trending. It shows mean reversion behavior throughout the data. 	

\begin{figure}
     \centering
     \begin{subfigure}[b]{0.45\textwidth}
         \centering
         \includegraphics[width=\textwidth]{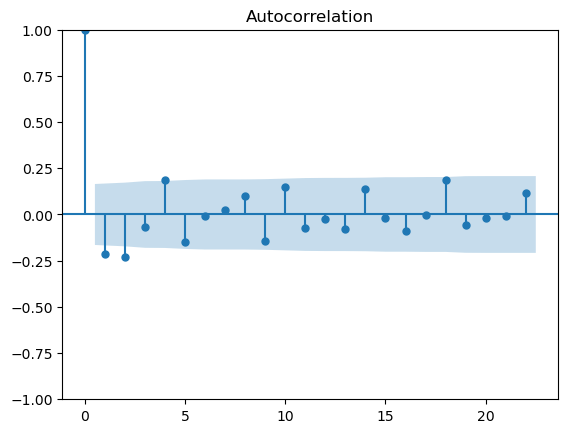}
         \vspace*{0.1cm}\centering{\textcolor{black}{lags}}
         \caption{ACF of first difference}
         \label{fig5}
     \end{subfigure}
     \hfill
     \begin{subfigure}[b]{0.45\textwidth}
         \centering
         \includegraphics[width=\textwidth]{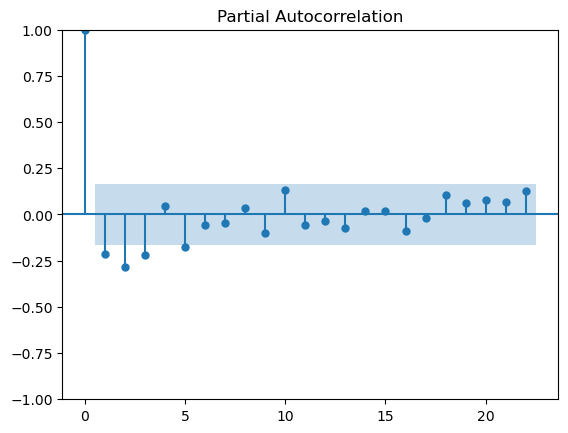}
         \vspace*{0.1cm}\centering{\textcolor{black}{lags}}
         \caption{PACF of first difference}
         \label{fig6}
     \end{subfigure}
     \hfill
        \caption{ACF and PACF of first difference}
\end{figure}

	\paragraph{}Also, we plot the autocorrelation for the first difference time series (Figure \ref{fig5}). Here after the first lag, we notice a sudden sharp decrease in ACF. Though there is very less difference between the second and the third lag. Almost all lags after that have ACF within the threshold value. But the second and third lag have values outside the threshold. Then we plot the Partial autocorrelation for the first difference time series (Figure \ref{fig6}). We notice that after the first lag, PACF shows a sudden decrease but it decays slowly after that. First four lags are outside the threshold. This shows that it is not a good model for the given time series.

		\begin{figure}[h]
		\centering
		\includegraphics[scale=0.93]{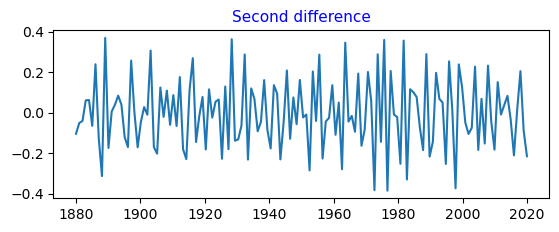}
              
            \vspace*{-0.1cm}\centering{\textcolor{black}{Year}}
		\caption{Second difference time series of the Global temperature data}
		\label{fig7}
	\end{figure}

\begin{figure}[h]
     \centering
     \begin{subfigure}[b]{0.45\textwidth}
         \centering
         \includegraphics[width=\textwidth]{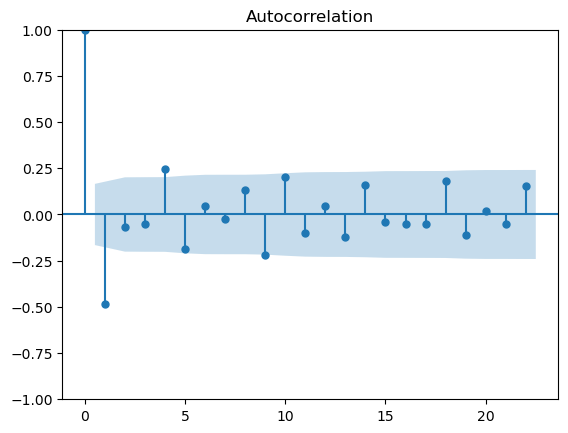}
         \vspace*{0.1cm}\centering{\textcolor{black}{lags}}
         \caption{ACF of second difference}
         \label{fig8}
     \end{subfigure}
     \hfill
     \begin{subfigure}[b]{0.45\textwidth}
         \centering
         \includegraphics[width=\textwidth]{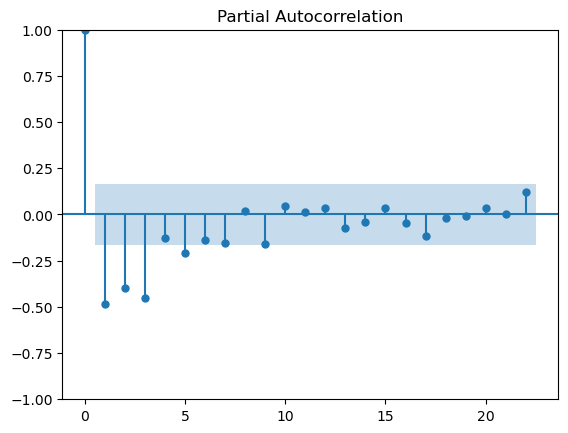}
         \vspace*{0.1cm}\centering{\textcolor{black}{lags}}
         \caption{PACF of second difference}
         \label{fig9}
     \end{subfigure}
     \hfill
        \caption{ACF and PACF of second difference}
\end{figure}

	 Figure \ref{fig7} is the second difference for the global temperature data. This also shows the mean reversion. 
	Further we plot the ACF and PACF for the second difference.
	Figure \ref{fig8} is the ACF for the second difference time series. Here the first two lags are much outside the threshold by magnitude. Lags after that are inside the threshold. From the behavior of the first two lags we can say that there is a possibility that the time series is over- differenced. Figure \ref{fig9} is the PACF graph for the second difference time series. First four lags here are outside the threshold and they have very high magnitude compared to the threshold. This confirms that the time series is overly differenced here and thus, this is not the best fitted model for the given Global data temperature.

	\subsection{Fitting AR$(1)$ Model}

	We try to fit the AR$(1)$ model in given data using autoarima function in python. The following is the summary of results:
	
	\includegraphics[scale=0.56]{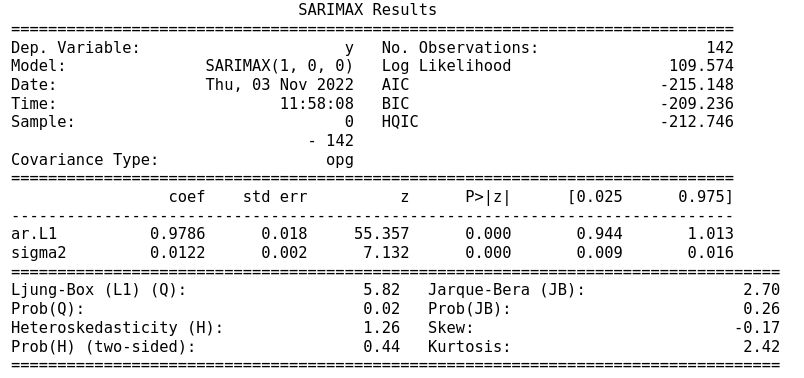}
	
	For the AR$(1)$ model we get the coefficients $\phi = 0.9786 $ and $\sigma = 0.0122$. This model fits with the skew $  -0.17$ and kurtosis $2.42$ . 
	
   \begin{figure}
       \centering
       \includegraphics[scale=0.5]{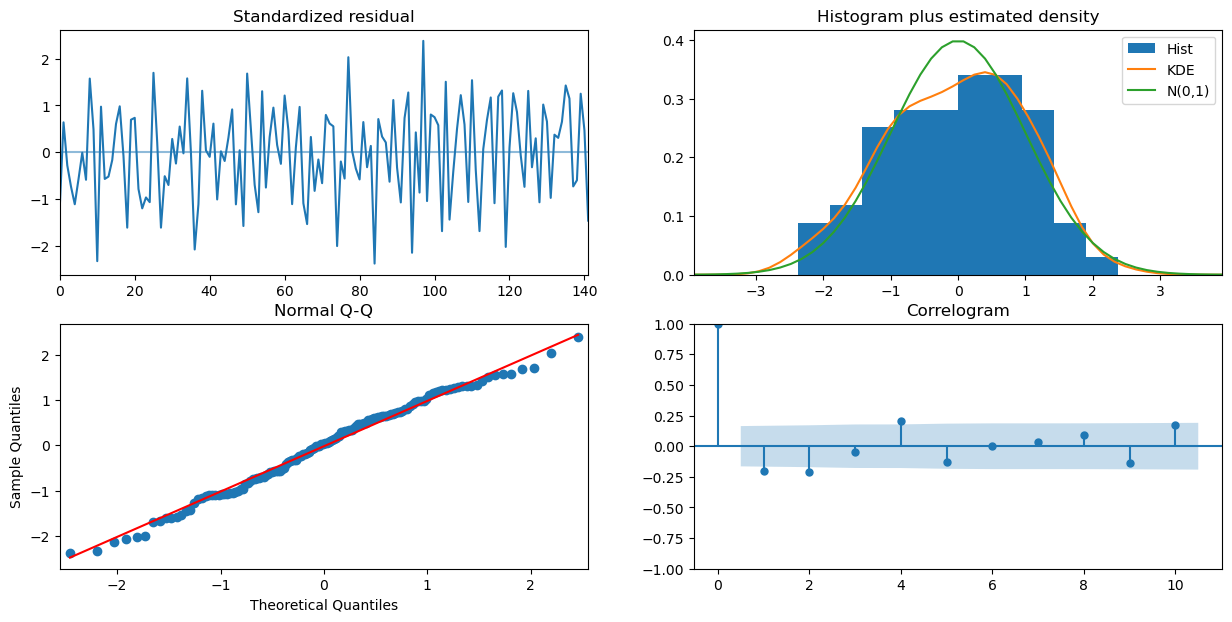}
       \caption{diagnostics for the AR$(1)$ model}
       \label{fig13AR}
   \end{figure}

Figure \ref{fig13AR} shows the diagnostics for the AR$(1)$ model. The first figure in this shows us standardized residuals.  Second figure shows the histogram for the data and Standard normal $(0,1)$ curve (in green) and the Kernel Density Estimation (KDE) graph (in orange) which smooths the given data. Third is the Normal quantile-quantile (Normal Q-Q) plot, where we can clearly see that most of the sample quantiles and theoretical quantiles fit the Normal distribution near the mean. But outside two standard deviations, it deviates from the reference line. Also, as we move away from first to second standard deviation, the data points start moving away from the reference line. Finally we have a correlogram. Here, we can see that this graph of ACF is very similar to the ACF plot of the first difference of the original time series, so this is not the best fitting model and we need to find a better model for this data.

		\begin{figure}[h]
		\centering
		\includegraphics[scale=0.5]{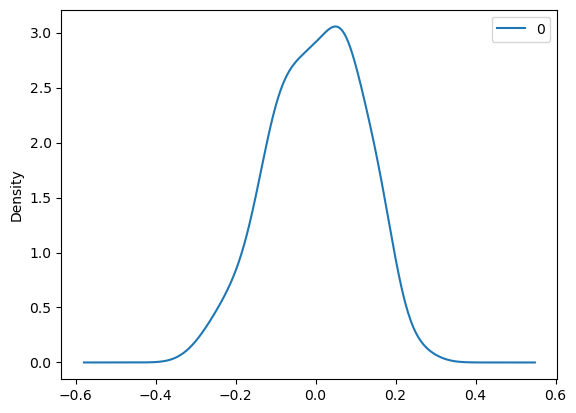}
            
            \vspace*{-0.25cm}\centering{\textcolor{black}{\tiny{Residue}}}
		\caption{Density of residuals for AR$(1)$}
		\label{fig15}
	\end{figure}

 Figure \ref{fig15} shows the density of residuals. Most of the residuals are near $0$. Also, from the residual analysis we can see that the maximum value we have for residuals for the AR$(1)$ model is $0.266279$.

	\subsection{Fitting ARMA$(1,1)$ Model}
	
	To get a better fit, we try to fit the ARMA$(1,1)$ model in given data using autoarima function in python. The following is the summary of results:
	
	\includegraphics[scale=0.56]{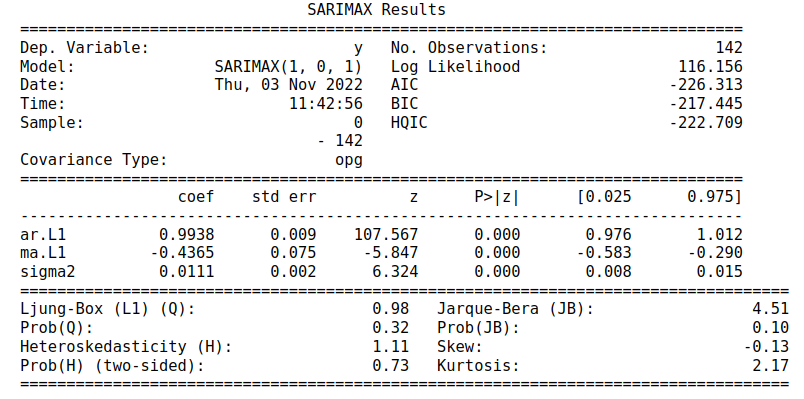}

	For the ARMA$(1,1)$ model we get the coefficients $\phi = 0.9938$, $\theta = -0.4365$ and $\sigma = 0.0111$. This model fits with the skew  $-0.13$ and kurtosis $2.17$ . 
	
	 \begin{figure}
       \centering
       \includegraphics[scale=0.5]{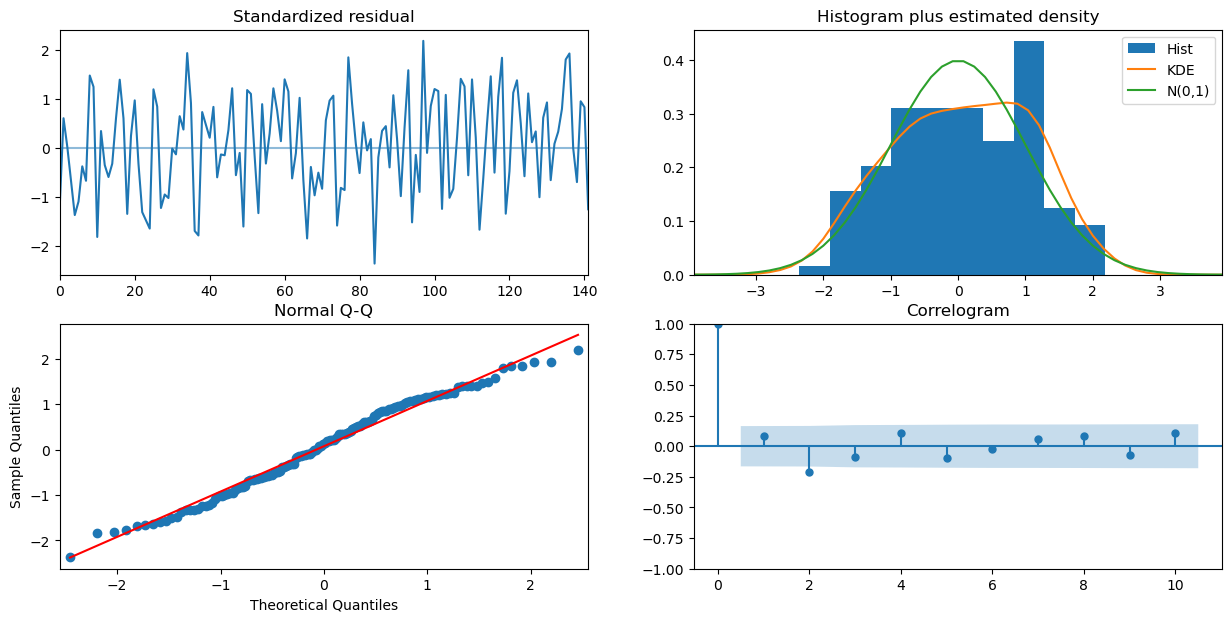}
       \caption{diagnostics for the ARMA$(1,1)$ model}
       \label{fig10ARMA}
   \end{figure}

	In Figure \ref{fig10ARMA} we have the diagnostics for the ARMA$(1,1)$ model. The first figure shows us standardized residuals. Comparing the histogram of ARMA$(1,1)$ to that of AR$(1)$, we can see that the KDE graph fits better and is closer to the standard normal graph. In the third graph, we can see that most of the sample quantiles and theoretical quantiles fit the Normal distribution near the mean. Here, between first and second standard deviation, very few points are away from the reference line. This plot clearly shows that as compared to the AR$(1)$ graph, ARMA$(1,1)$ has a better fit to given data. The correlogram here is different from the ACF plot of the original time series, and the second lag is inside the threshold. 

 \begin{figure}[h]
		\centering
		\includegraphics[scale=0.5]{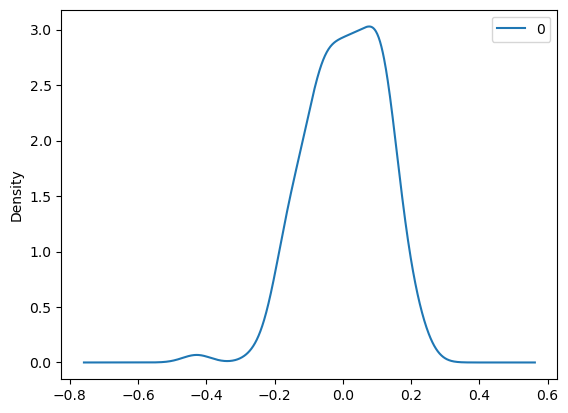}
              
            \vspace*{-0.25cm}\centering{\textcolor{black}{\tiny{Residue}}}
		\caption{Density of residuals for ARMA$(1,1)$}
		\label{fig12}
	\end{figure}

   Figure \ref{fig12} shows the density of residuals. Most of the residuals are near $0$. From the analysis of residuals of ARMA$(1,1)$ we can see that the maximum value we have for residuals is $0.232144$, which is less than the maximum residual value for AR$(1)$ model. Thus ARMA$(1,1)$ is much better model than AR$(1)$.

       	\subsection{Fitting ARIMA$(1,1,1)$ and ARIMA $(1,2,0)$ Models}
	In general situation we are  aware that ARIMA $(p,d,q)$ models fit better that ARMA models. So, we try to fit ARIMA models with the difference $1$ and $2$ in the GISTEMP data.. For $d=1$, simulation shows us that ARIMA$(1,1,1)$ is the best model and for $d=2$, we get ARIMA$(1,2,0)$ as the best fitting model.
	
	Similar to previous fittings, we try to fit the ARIMA$(1,1,1)$ model using autoarima function in python. The following is the summary of results:

       \includegraphics[scale=0.56]{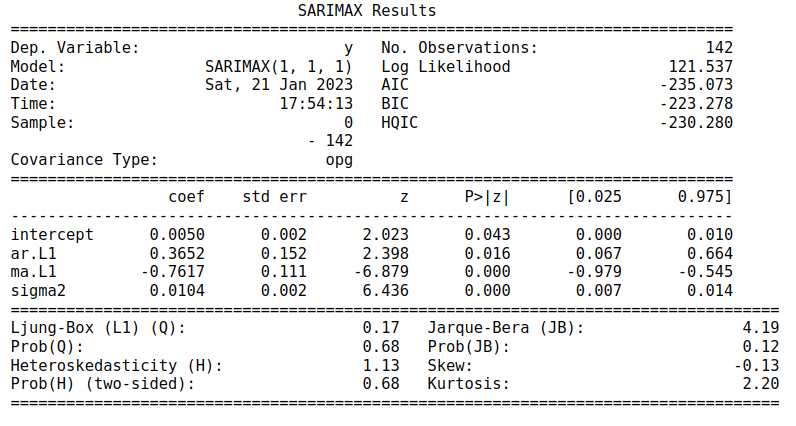}

    For the ARIMA$(1,1,1)$ model the coefficients $\phi = 0.3652$, $\theta = -0.7617$ and $\sigma = 0.0104$. This model fits with the skew  $-0.13$ and kurtosis $2.20$ . 
	
	 \begin{figure}
       \centering
       \includegraphics[scale=0.33]{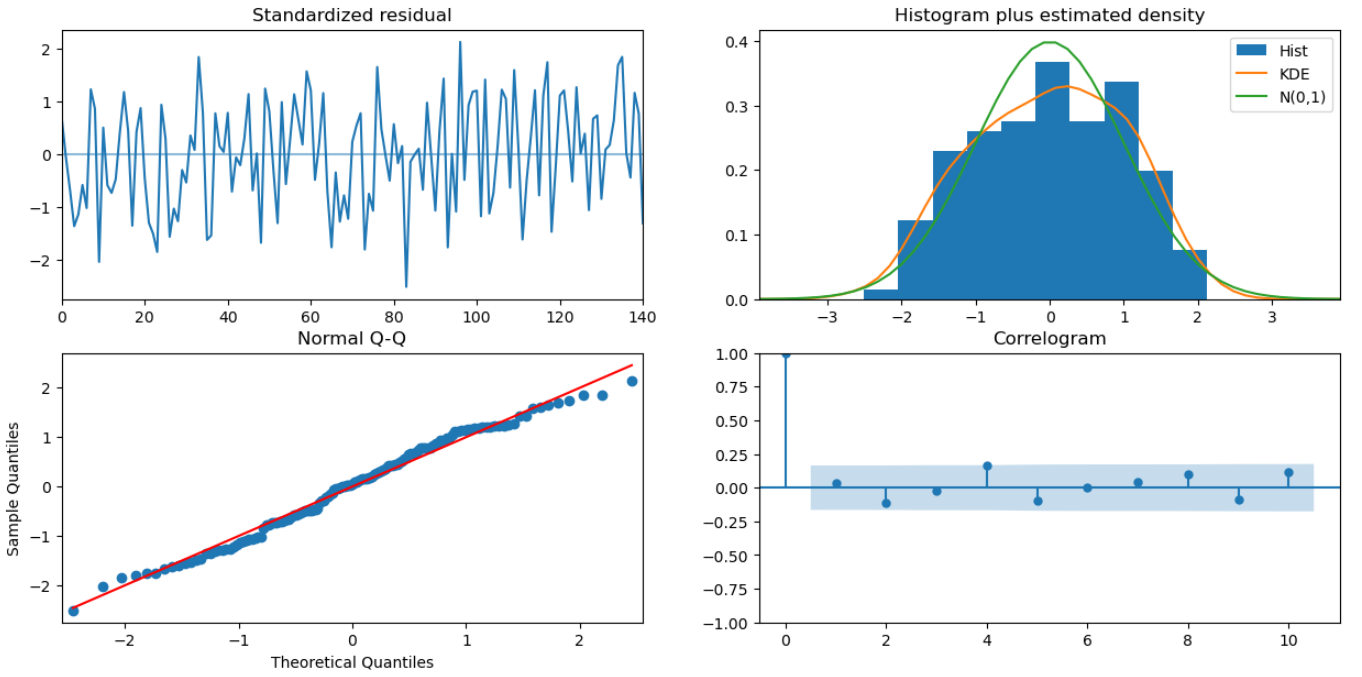}
       \caption{diagnostics for the ARMA$(1,1,1)$ model}
       \label{fig16ARIMA111}
   \end{figure}

In Figure \ref{fig16ARIMA111} we have the diagnostics for the ARIMA$(1,1,1)$ model. The first figure shows us standardized residuals. Comparing the histogram of ARMA$(1,1)$ to ARIMA$(1,1,1)$, we can see that the KDE graph fits better in ARMA$(1,1)$ and is closer to the standard normal graph. The Normal Q-Q graph and the correlogram for ARIMA$(1,1,1)$ does not look much different than that of ARMA$(1,1)$.

\begin{figure}[h]
		\centering
		\includegraphics[scale=0.35]{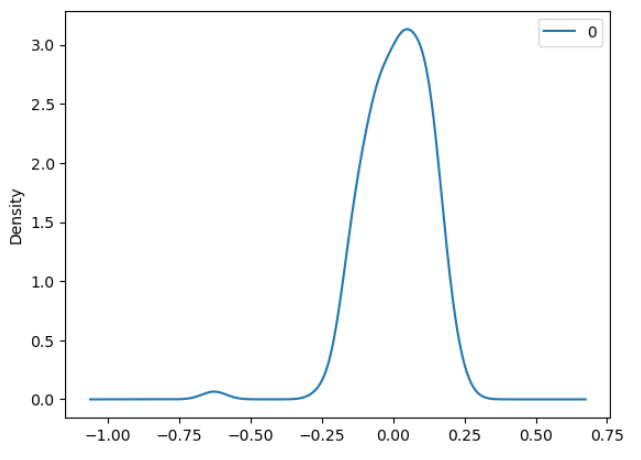}
              
            \vspace*{-0.25cm}\centering{\textcolor{black}{\tiny{Residue}}}
		\caption{Density of residuals for ARIMA$(1,1,1)$}
		\label{Fig16c}
	\end{figure}

  Figure \ref{Fig16c} shows the density of residuals. Most of the residuals are near $0.1$. From the analysis of residuals of ARIMA$(1,1,1)$ we can see that the maximum value we have for residuals is $0.240199$. Which is higher than the maximum residual value for ARMA$(1,1)$ model. Thus ARMA$(1,1)$ is better model than ARIMA$(1,1,1)$.

   Further, we try to fit an ARIMA model with difference $d=2$. The autoarima function in python, gives us that ARIMA$(1,2,0)$ is the best fitting model for given data for $p \leq 1 $ and $q\leq 1$. When we fit ARIMA$(1,2,0)$ we get following results:

   \includegraphics[scale=0.56]{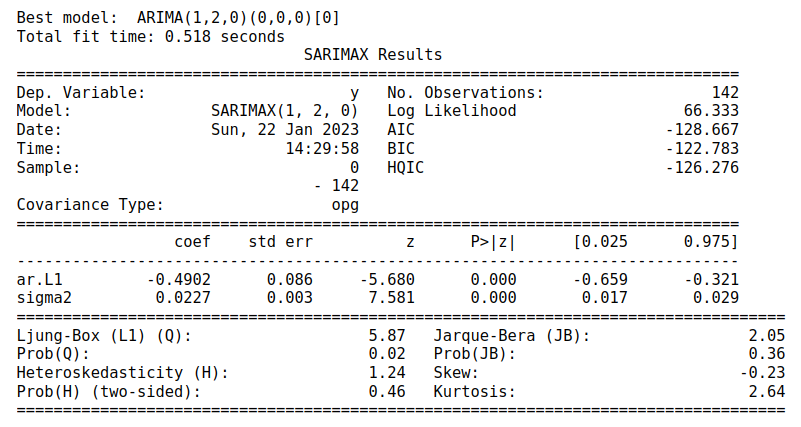}

   For the ARIMA$(1,2,0)$ model the coefficients $\phi = -0.4902$, $\theta = 0$ and $\sigma = 0.0227$. This model fits with the skew  $-0.23$ and kurtosis $2.64$ .

 \begin{figure}
       \centering
       \includegraphics[scale=0.33]{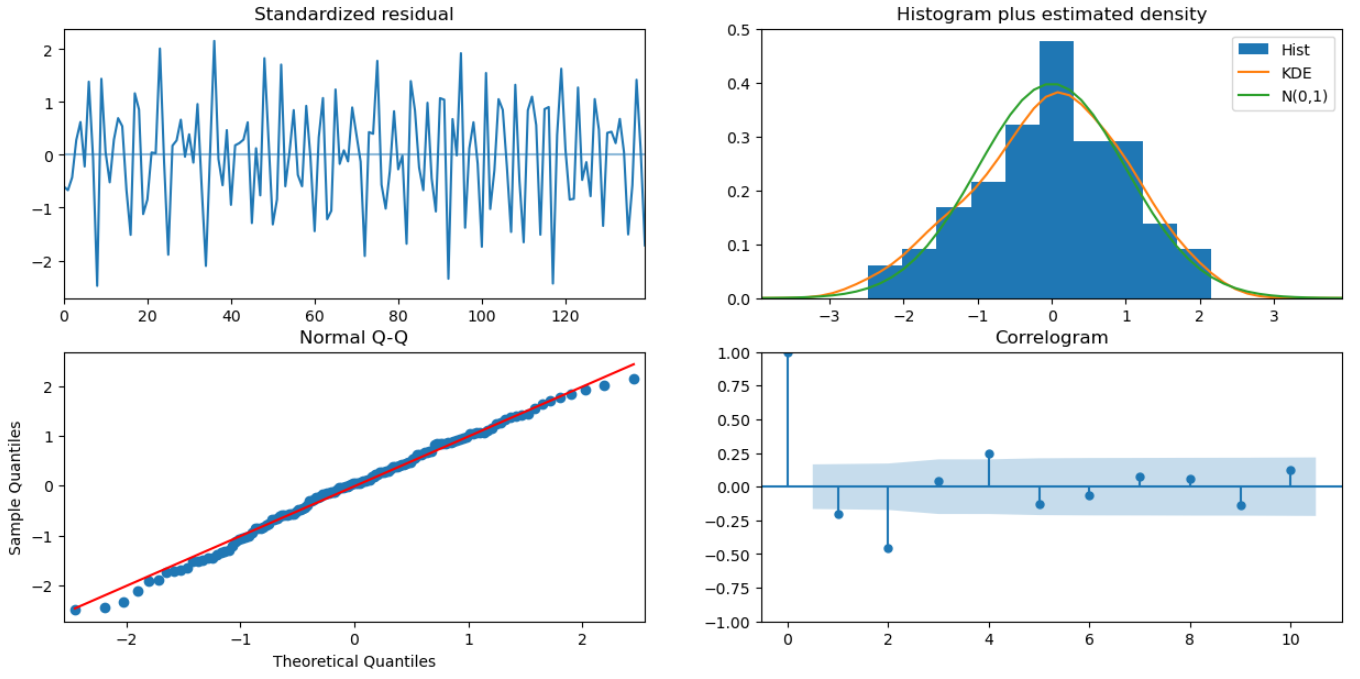}
       \caption{diagnostics for the ARMA$(1,2,0)$ model}
       \label{fig17ARIMA120}
   \end{figure}

In Figure \ref{fig17ARIMA120} we have the diagnostics for the ARIMA$(1,2,0)$ model. The first figure shows us standardized residuals. Comparing the histogram of ARMA$(1,1)$ and ARIMA$(1,2,0)$, we can see that the KDE graph here does not have better fit. The Normal Q-Q graph for ARIMA$(1,2,0)$ does not look much different than that of ARMA$(1,1)$. But in correlogram we notice different behaviour of ACF. Here ACF drops suddenly after first lag and second and third are outside the threshold. After that most of the lags are inside,  but they clearly are not converging towards zero. This clearly shows that this model is not a good fit.

\begin{figure}[h]
		\centering
		\includegraphics[scale=0.35]{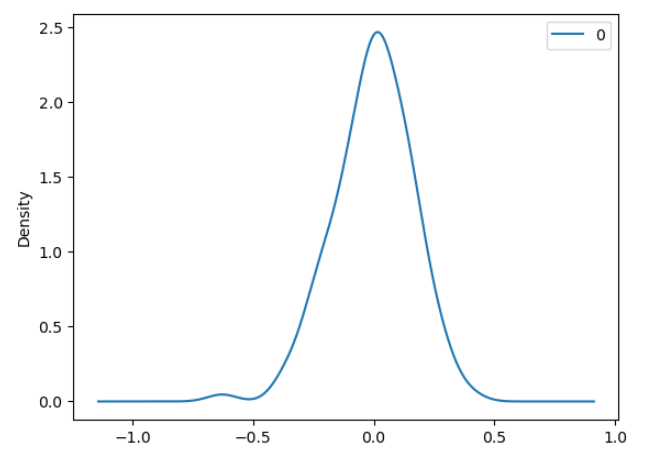}
              
            \vspace*{-0.25cm}\centering{\textcolor{black}{\tiny{Residue}}}
		\caption{Density of residuals for ARIMA$(1,2,0)$}
		\label{Fig17c}
	\end{figure}

Further, the density of residuals in Figure \ref{Fig17c} reassures the result. Most of the residuals for ARIMA$(1,2,0)$ are near $0$. From the analysis of residuals we can see that the maximum value we have for residuals is $0.398333$. Which is higher compared to the maximum residual value for ARMA$(1,1)$ model. 

Hence, from the analysis of best fitting models for difference $d=1$ and $d=2$, i.e. ARIMA$(1,1,1)$ and ARIMA$(1,2,0)$ models, we can see that ARMA$(1,1)$ model is a better fit than those models.

	\section{Conclusion:}

	Our new development of ARMA $(1,1)$  based on uncertainty in the global mean anomaly due to uncertainties in the land and sea surface was analyzed and compared with AR$(1)$, ARIMA$(1,1,1)$ and ARIMA$(1,2,0)$. In general theory, ARIMA $(p,d,q)$ models are considered to have a better fit than that of ARMA$(p,q)$ models. But for the Goddard Institute for Space Studies Surface Temperature (GISTEMP) Data from 1880 to the present we found different result. Here the ARMA$(1,1)$ model is a better fitting model than the ARIMA$(1,1,1)$ and ARIMA$(1,2,0)$. Also, from simulations it was evident that ARIMA$(1,2,0)$ was a better fitting model than ARIMA$(1,2,1)$, so it concludes the fact that ARMA$(1,1)$ is a better model for given data than ARIMA$(1,d,1)$ for $d=1$ and $d=2$.

 The forecast for ARMA$(1,1)$ is unbiased and forecast error variance increases without bounds as the lead time increases. For nonstationary series, when we forecast far into the future, we have a lot of uncertainty about the forecast. Moreover, From the Normal Q-Q plot, correlograms, and the KDE plot, we can say that the ARMA$(1,1)$ model is a better fit. Furthermore, residual analysis reassures the result.

 These results are true only for the Goddard Institute for Space Studies Surface Temperature (GISTEMP) Data and authors want to specify that the analysis in this article and results may or may not hold true for other data.

\textbf{Data availability statement}\\
The original contributions of data were taken from \url {https://data.giss.nasa.gov/gistemp/graphs/},and the simulation was done by Python, further inquiries can be directed to the corresponding author.

\textbf{Author contributions }\\
All authors listed have made a substantial, and intellectual effort for this research.

\textbf{Acknowledgments  } \\
The authors would like to thank Dr. Sang, Associate Prof., Dept. of Mathematics, the University of Mississippi for teaching the course on Time Series Analysis where we learned and got the statistical idea about the Time series model.

\textbf{Conflict of interest  }
The authors declare that the research was conducted in the absence of any commercial or financial relationships that could be construed as a potential conflict of interest.

	\end{document}